\documentclass[twocolumn,superscriptaddress]{revtex4-2} 
\usepackage[T1]{fontenc}
\usepackage{times}
\usepackage{graphicx}
\usepackage{color}
\usepackage{booktabs}
\newcommand{\cl}{\textcolor{black}}

\usepackage{amsfonts, amsmath,amssymb, bm, graphicx}
\usepackage{siunitx}

\usepackage{amssymb}
\usepackage[english]{babel}
\usepackage{lipsum}

\newcommand{\figref}[2]{\hyperref[#1]{\ref{#1}(#2)}}
\newcommand{\figrefsub}[3]{\hyperref[#1]{\ref{#1}(#2)#3}}

\usepackage{physics}
\usepackage{lipsum}
\usepackage{changes}

\usepackage[colorlinks=true,allcolors=blue]{hyperref}
\usepackage{footmisc}

\usepackage{upgreek}

\usepackage{soul}

\makeatletter
\let\ORIbbl@fixname\bbl@fixname
\def\bbl@fixname#1{%
  \@ifundefined{languagealias@\expandafter\string#1}
    {\ORIbbl@fixname#1}
    {\edef\languagename{\@nameuse{languagealias@#1}}}%
}
\newcommand{\definelanguagealias}[2]{%
  \@namedef{languagealias@#1}{#2}%
}
\makeatother

\definelanguagealias{en}{english}

\begin{document}

{
\makeatletter
\def\frontmatter@thefootnote{%
 \altaffilletter@sw{\@fnsymbol}{\@fnsymbol}{\csname c@\@mpfn\endcsname}%
}%

\makeatother
\title{Nontrivial Aharonov-Bohm effect and alternating dispersion of magnons in cone-state ferromagnetic rings}

\author{Vera Uzunova}
\affiliation{Institute of Theoretical Physics, Faculty of Physics, University of Warsaw, ul. Pasteura 5, 02-093 Warszawa, Poland}
\affiliation{Institute of Physics of the National Academy of Sciences of Ukraine, 46 Nauky Ave., 03039 Kyiv, Ukraine}

\author{Lukas K\"orber}
\affiliation{Helmholtz-Zentrum Dresden - Rossendorf, Institut f\"ur Ionenstrahlphysik und Materialforschung, D-01328 Dresden, Germany}
\affiliation{Fakultät Physik, Technische Universit\"at Dresden, D-01062 Dresden, Germany}

\author{Agapi Kavvadia}
\affiliation{Helmholtz-Zentrum Dresden - Rossendorf, Institut f\"ur Ionenstrahlphysik und Materialforschung, D-01328 Dresden, Germany}
\affiliation{Fakultät Physik, Technische Universit\"at Dresden, D-01062 Dresden, Germany}

\author{Gwendolyn Quasebarth}
\affiliation{Helmholtz-Zentrum Dresden - Rossendorf, Institut f\"ur Ionenstrahlphysik und Materialforschung, D-01328 Dresden, Germany}
\affiliation{Fakultät Physik, Technische Universit\"at Dresden, D-01062 Dresden, Germany}

\author{Helmut Schultheiss}
\affiliation{Helmholtz-Zentrum Dresden - Rossendorf, Institut f\"ur Ionenstrahlphysik und Materialforschung, D-01328 Dresden, Germany}
\affiliation{Fakultät Physik, Technische Universit\"at Dresden, D-01062 Dresden, Germany}

\author{Attila K\'{a}kay}
\affiliation{Helmholtz-Zentrum Dresden - Rossendorf, Institut f\"ur Ionenstrahlphysik und Materialforschung, D-01328 Dresden, Germany}

\author{Boris Ivanov}
\affiliation{Institute of Magnetism, National Academy of Sciences of Ukraine, 03142 Kiev, Ukraine}
\affiliation{Institute for Molecules and Materials Radboud University Nijmegen
Heyendaalseweg 135   6525 AJ Nijmegen The Netherlands}

\date{\today}
\begin{abstract}    
 Soft magnetic dots in the form of thin rings have unique topological properties. They can be in a vortex state with no vortex core. Here, we study the magnon modes of such systems both analytically and numerically. In an external magnetic field, magnetic rings are characterized by easy-cone magnetization and shows a giant splitting of doublets for modes with the opposite value of the azimuthal mode quantum number.  The effect of the splitting can be refereed as a magnon analog of the topology-induced Aharonov-Bohm effect. For this we develop an analytical theory to describe the non-monotonic dependence of the mode frequencies on the azimuthal mode number, influenced by the balance between the local exchange and non-local dipole interactions.
\end{abstract}

\maketitle

\section{Introduction}

One of the most attractive features of magnetic nanostructures is
the presence of topological defects (such as vortices, skyrmions,
monopoles (Bloch points), domain walls, and chiral bobbers) as their
magnetic equilibrium configurations, sometimes even the ground
state. Among these spin textures, magnetic vortices, a spontaneously forming
ground state in (sub)micron soft-magnetic discs have attracted
great attention during the past two decades \cite{guslienkoMagneticVortexCore2006,zhangDeterministicReversalSingle2020}. The magnetic vortex consists
of a curling in-plane magnetization with closed magnetic flux around a
the nanoscale region known as the vortex core with out-of-plane
magnetization. The vortex state has been extensively studied,
partly due to its rich excitation spectrum \cite{ivanovMagnonModesMagnonvortex1998,shekaAmplitudesMagnonScattering2004,ivanovHighFrequencyModes2005,buessExcitationsNegativeDispersion2005,taurelCompleteMappingSpinwave2016} and the possibility of
the usage in magnonic or spintronic devices, for example, as storage units \cite{shinjoMagneticVortexCore2000,vanwaeyenbergeMagneticVortexCore2006,hertelUltrafastNanomagneticToggle2007},  nano-oscillators \cite{guslienkoSpinTorqueCritical2011,mistralCurrentDrivenVortexOscillations2008,pribiagMagneticVortexOscillator2007,ruotoloPhaselockingMagneticVortices2009,slukaSpintorqueinducedDynamicsFinesplit2015}, or even for neuromorphic computing \cite{korberPatternRecognitionReciprocal2023}. 

Spectrum of the
vortex-state particle includes the gyrotropic mode with low
(sub-gigahertz) frequency $\omega_{g}$, corresponding to the vortex
core precession around the disc center, and a system of higher modes.
These modes with the frequencies $\omega_{m,n}$ are denoted by two
integers, the number of radial nodes in the out-of-plane component
of the dynamical magnetization (radial mode number) $n\geq 1$ and
the azimuthal mode number $m$, which determines the angular
dependence on this component. These modes form doublets with the
splitting $\Delta \omega_{n,m}\equiv
\omega_{n,|m|}-\omega_{n,-|m|}$. 

\cl{The model describing magnon
dynamics on a vortex background that takes into account non-local magnetostatic dipole interaction leads to integrodifferential Landau-Lifshitz equations. In general case, this problem cannot be solved analytically.}
Current successes in the
understanding of magnon modes for the vortex-state soft-magnetic
particles are partly based on past investigations of these modes
for easy-plane two-dimensional (2d) models with local interactions
only, see \cite{ivanovMagnonModesMagnonvortex1998}. Indeed, in case of vortex-state disk the gyroscopic mode is present for both
models, with the same sense of the vortex precessional mode, which
can be described on the ground of the Thiele equation for vortex core
coordinate. For both models, the frequencies of modes are determined by small
parameters of the problems, which are $(a/R)^2$ for the 2d local
model and $L/R$ for soft magnetic discs, where $R$ is the radius of
the disc, $a$ is the lattice constant and $L$ is the disc thickness.
The higher modes were found to form a set of doublets with given
value of $n$ and $m=\pm |m|$, their mean frequencies are
proportional to the square root of these small parameters, $(a/R)$ and
$\sqrt{L/R}$ for both models. It was found that the interaction with
the vortex core is significant for higher translation modes with
$m=\pm 1$, leading to doublet splitting $\Delta \omega_{n,1}$ of the
order of $\omega_{g}$. Surprisingly the ratio $\Delta
\omega_{1,1}/\omega_{g}\cong 3.5$ is the universal value for these
two quite different models. The dynamics of vortex core with
accounting for higher modes leads to higher order Thiele equation,
first proposed for a local model, see \cite{ivanovMagnonModesMagnonvortex1998}, and
confirmed by numerical simulations for vortex state soft-magnetic
dots \cite{khvalkovskiyVortexOscillationsInduced2009,IvanovAvanesyan2010}.

For a soft ferromagnetic disk, the leading contribution in magnon
frequencies is provided by the strong dipolar interaction of magneto-static
origin \cite{buessPulsedPrecessionalMotion2003,buessExcitationsNegativeDispersion2005}. Near the vortex core, the magnon wave functions
strongly decrease as $r^{|m-1|}$, i.e., magnon for modes with
azimuthal number $|m|>1$ are delocalized from the core region. This asymptotic behavior is valid for the easy-plane local model and the soft ferromagnet. Thus
the interaction of these magnons with the vortex core, particularly the doublet splitting, is expected to be weak for these
modes. Indeed, it is the case for standard in-plane vortex. Nevertheless, one exception was found for the cone vortex state, created by
an external magnetic field applied along the hard axis. In this case, the magnetization far from the core is tilted relative to the disc plane. For the cone vortex state, the significant splitting, linear over the
value of the magnetic field appears, as discussed in Ref.~\cite{ivanovMagnonModesCircular2002}. 
Other works by Dugaev and Bruno~\cite{dugaevBerryPhaseMagnons2005,brunoBerryPhaseTopology2006} have discussed such a splitting in vortex-state ferromagnetic rings without a core but tilted magnetization. Similar splitting was also found in helical-state magnetic nanotubes \cite{salazar-cardonaNonreciprocitySpinWaves2021}. In these works, the dominant role of the local exchange interaction was highlighted, with Ref.~\cite{dugaevBerryPhaseMagnons2005} connecting the frequency splitting to a geometrical phase arising due to the nontrivial Berry connection in inhomogeneous magnetic equilibria. These recent works, however, have only discussed the role of local interactions.

\cl{Here, we show that giant doublet splitting is a generic feature related to the nontrivial topology of the vortex cone state, by developing a model including both local exchange and non-local dipole interactions. }The topology of magnetic textures gives rise to a variety of fundamentally new effects, such as skew and rainbow scattering the magnon modes in the presence of a single skyrmion in a ferromagnet \cite{iwasakiTheoryMagnonskyrmionScattering2014} and a chiral magnet \cite{schutteMagnonskyrmionScatteringChiral2014}; it also results in a finite gyro coupling force in the Thiele equation \cite{IvanovSheka}. Soft magnetic dots with easy-cone magnetization possess unique topological properties, absent for other cases, either skyrmions or planar vortices. To this end, we demonstrate how the topology of a vortex gives rise to the magnon analog of the Aharonov-Bohm effect that results in giant doublet splitting. We also describe the nontrivial dependence of the mean magnon frequency on the azimuthal number. The studies were conducted analytically and numerically, with good agreement between the two approaches. 

Therefore, in Sec.~\ref{sec:topology}, we discuss the general topological features of the cone vortex state in magnetic nanodots and nanorings, which ultimately give rise to the topology-induced Aharonov-Bohm effect. In Sec.~\ref{sec:analytical_model}, we present an approximate analytical model of the magnon spectrum about the cone-vortex equilibrium state, including both local and non-local interactions. In order to disentangle the giant doublet splitting in the cone state from the splitting of the $m=\pm 1$ doublet, which appears already due to the presence of a vortex core, our theory is applied to a ferromagnetic ring without a core. In Sec.~\ref{sec:validation}, our model is compared with exact numerical solutions of the linearized equation of motion that governs the magnon spectrum, providing good agreement. Our approximate analytical theory describes the magnon dispersion in the cone-vortex state and directly attributes its nontrivial dependence on the azimuthal mode number $m$ to the interplay between local and non-local interactions. Finally, we demonstrate how both of these interactions contribute to the topology-induced giant doublet split. Besides providing an analytical model for magnon dynamics in (cone)-vortex-state nanodots and rings, our paper contributes to the fundamental understanding of topological effects in magnon dynamics and magnetism, in general. \textcolor{black}{We also discuss consequences on potential applications based on cone-state vortex rings.}

\section{ {Vortex structure and topology-induced Aharonov-Bohm effect.}}\label{sec:topology}

\cl{We consider a soft magnetic particle in a form of a thin ring with an inner radius $R_\mathrm{in}$, outer radius $R_\mathrm{out}$, and thickness $L$.
The inner and outer radii are much larger than the exchange length $l_0$ of the ferromagnetic material. The magnetization vector is denoted $M(\mathbf{r})$, normalized magnetization $m(\mathbf{r})=M(\mathbf{r})/M_s$, where $M_\mathrm{s}$ is the saturation magnetization. An external magnetic field $\mathbf{H}_0$ is applied along the axis of the ring. In what follows, we use the cylindrical basis $(\bm{e}_r,\bm{e}_\chi,\bm{e}_z)$ with the $\bm{e}_z$ along the axis of the particle.}

\cl{
The ground state of the thin magnetic particle in a shape of a disk or a ring is the vortex ground state, in which the magnetization circulates around the center region to close the magnetic flux. The vortex state was obtained as a solution of the Landau-Lifshitz equations $\partial\mathbf{M}/\partial t=\gamma[\mathbf{M}\times (\delta W/ \delta \mathbf{M})]$. Here $\gamma$ is the gyromagnetic ratio, and $(\delta W/ \delta \mathbf{M})$ denotes the variation of energy functional of the ferromagnet. We will discuss the specific form of the functional $W=W[\mathbf{M}]$ in the next Section. Here we just note, that only the local exchange interaction contributes to the vortex solution. The contribution of the magnetostatic dipole interaction to Landau-Lifshitz equations is proportional to $\div\mathbf{M}$, which for the vortex ground state is zero.}

Within the standard parametrization of the
normalized magnetization  $\bm{m}=(\sin\theta\cos\phi,\sin\theta\sin\phi,\cos\theta)$ the vortex structure can be described as $\phi=q\chi+\pi/2,\;\theta=\theta_0(r)$, with $q$ being an integer, see Ref.~\cite{ivanovMagnonModesCircular2002}.
 Here the function $\theta_0(r)$ satisfies the ordinary differential equation
  \begin{eqnarray}\label{}
\frac{d^2\theta_0}{dr^2}+\frac{d\theta_0}{rdr}+\sin\theta_0\cos\theta_0\left(\frac{1}{l_0^2}-\frac{q^2}{r^2}\right)=\frac{h\sin\theta_0}{l_0^2}
\end{eqnarray}
with the natural boundary conditions 
$\sin\theta_0(0)=0$ and
$\cos\theta_0(\infty)=h$. \cl{Here, $h$ is the dimensionless strength of the external field: $\bm{H}_0=4\pi M_s h\bm{e}_z$.}

The set of possible values of $\bm{m}$ in the equilibrium state form the
order parameter space. From a topological point of view, the vortex state is described by the group of continuous mappings of the coordinate into the order parameter space. This, in particular, is the relative homotopy group $\pi_2(S^2,S^1)$ of mappings of the two-dimensional
sphere $S^2$ with the non-contractible boundary $S^1$, because of the non-equivalence of $\bm{m}$ and
$-\bm{m}$ magnetization.

\begin{figure}[h]
\includegraphics{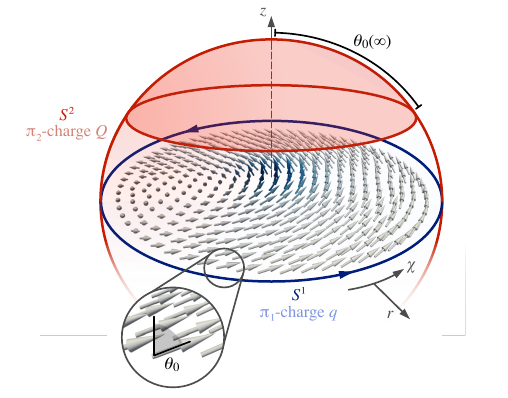}
\caption{Structure of a cone-state magnetic vortex with angle $\theta_0(r)$ to the $z$ axis. A cylindrical basis $(r,\chi,z)$ is defined as shown. The winding number of the magnetization along the closed loop $S^1$ defines a $\pi_1$-topological charge (vorticity). The mapping of a sphere $S^2$ to the order parameter space defines a $\pi_2$-topological charge (also referred to as Skyrmion number).
\label{Sphere}}
\end{figure}

The vortex is characterized by two topological invariants. The
$\pi_1$-topological charge is known as vorticity
\begin{equation}
    q=\frac{1}{2\pi}\int\limits_{S^1}{\partial_\chi\phi} \,\mathrm{d} \chi 
\end{equation}
counting the
accumulation of phase $\phi$ along a remote closed path $S^1$ in
coordinate space embracing the vortex center.

The mapping of the $S^2$ sphere to the vortex configuration defines the $\pi_2$-topological charge, also known as Skyrmion number, as
\cl{
\begin{equation}\label{pi2}
    Q=\frac{1}{4\pi}\int\limits_{S^2} \mathcal{Q}r{d} r{d}\chi
\end{equation}}
with the topological-charge density,
\begin{equation}
    \mathcal{Q} = \bm{m}
\cdot(\partial_r \bm{m}\times
\partial_\chi \bm{m}),
\end{equation}
playing the role of a gyrocoupling  density \cite{shekaAmplitudesMagnonScattering2004}. In terms of the spherical magnetization angles, it can be given as 
\begin{eqnarray}\label{}
\mathcal{Q}=\frac{q\sin\theta_0}{r}\left(\frac{d\theta_0}{d r}\right).
\end{eqnarray}
\color{black}
The value of $Q$ is directly determined by the boundary conditions
imposed on the polar angle of magnetization
\cl{
\begin{eqnarray}\label{}
Q =\frac{q}{2}\int\limits_{\theta_0(0)}^{\theta_0(\infty)}\sin\theta_0{d}\theta_0
=\frac{q}{2}\left[\cos\theta_0(0)-\cos\theta_0(\infty)\right].
\end{eqnarray}}

The boundary condition in the center plays the role of polarization
$p=\cos\theta_0(0)$ which takes integer values $\pm 1$ depending on
whether the direction of $\bm{m}$ coincides with the direction
of $z$ axis or is opposite to it.

In the case of a zero external field, the configuration of an planar vortex takes place. Then $\theta_0(\infty)=\pi/2$ and $Q={pq}/{2}$, which corresponds to the mapping on the upper or lower hemispheres depending
on the sign of polarization. In this case, the planar vortices with opposite
polarization ($p=\pm 1$) are energetically degenerate. Applying the external field the planar vortex transforms into
the cone-state vortex, in which the vector $\bm{m}$  can lie inside or outside the cone with
opening angle $2\theta_0(\infty)=2\arccos h$, see Fig.~\ref{Sphere}. Thus there are vortices of two different types, corresponding to different values of polarization $p=\pm 1$. They have different energy, the vortex with polarization along the magnetic field has lower energy and is called light vortex. The vortex with the opposite value of polarization has higher energy and is called the heavy one \cite{ivanovMagnonModesCircular2002}.
For light vortex with $p=1$ mapping of $S^2$ is an upper spherical segment
shown on figure by red color, for heavy vortex with $p=-1$ it is the remaining
part of the sphere shown by white color. The topological charge
$Q=q(p-h)/2$ implies in general transcendent values are
responsible for unique topological properties of easy-cone vortices
regardless of their specific nature.

The topological features of the system also show themselves in the fact that in the linearized Landau-Lifshitz equations describing the magnon dynamics contain a formal magnetic field  $\bm{B}=\mathrm{rot}\,\bm{A}=
\mathcal{Q}\bm{e}_z$, see Ref.~\cite{shekaAmplitudesMagnonScattering2004}. This field acts on magnons in a way the usual magnetic field influences electrons.  For example, in the local model the
linearized equations can be rewritten in the form
of a generalized Schr\"{o}dinger equation for a complex magnon wave
function, \cite{shekaAmplitudesMagnonScattering2004}, 
\begin{equation}\label{GShr}
    i\partial_t
\psi=(-i\nabla-\bm{A})^2+U)\psi+W\psi^*,
\end{equation}
which contains the vector potential $\bm{A}$ in a standard way. The total magnetic flux passing through any
closed surface embracing the vortex
$\Phi=\int\bm{B}d\bm{s}=\oint\bm{A}d\bm{l}$ is
exactly
\begin{eqnarray}\label{Phi}
\Phi=4\pi Q=q[p-h]\Phi_0.
\end{eqnarray}
with the flux quantum $\Phi_0=2\pi$.

In the original Aharonov-Bohm effect, this total flux determines the
interference between charged particles traveling along different paths
around a long (formally infinite) solenoid \cite{aharonovSignificanceElectromagneticPotentials1959}. In particular, the magnetic flux is
concentrated inside the solenoid while the movement of the particles takes
place outside the magnetic field region where
$\bm{B}=\textrm{rot}\,\bm{A}=0$. The Hamiltonian of the system
explicitly contains the vector potential and the resulting Schr\"{o}dinger equation
has a general solution  expressed in terms of Bessel function of
transcendent index $J_{|m+\alpha|}$ where $\alpha=-\Phi_s/ \Phi_0 $ is
the phase difference acquired by particle while traveling around
solenoid, with \cl{ $ \Phi_0= 2\pi c\hbar/e $ } being the flux quantum. When the total flux
inside the solenoid $\Phi_s$ is a multiple of flux quantum $
\Phi_0$ the index of Bessel function is integer-valued and the interference
is constructive (the phase shift depend only on difference between the
traveled paths). Non-integer $\alpha$ means that wave function
acquire additional phase factor and the interference pattern shifts.

Even though magnons are not charged particles, the nontrivial
topological charge of the vortex core influences the magnon
excitations even when the they do not enter core region. If the
total flux $\Phi$ defined by Eq.~(\ref{Phi}) is not an integer multiple
of the flux quantum the wave functions of the magnons become dependent
on sign of the azimuthal number $m$ and splitting for modes with
different sense of gyration appears. This effect is general in the
sense that cone state of a magnetic particle can be created not only by a magnetic field. In fact, the same effects are obtained in a magnet with the complex crystallographic anisotropy, including second and fourth order contributions, so that the ground state is realised by the cone state \cite{uzunovaMagnonModesMagnetic2019}.

\cl{For comparison we can mention splitting that occurs due to the magnetostatic dipole interaction, but it is not associated with the Aharonov-Bohm effect \cite{APL116}. In the case of topologically induced splitting, which we consider, the effect exists even for the pure local model, that does not take into account the magnetostatic interaction.}

\cl{The topological splitting of magnons is a unique property of the cone-state vortex regardless of its nature, \cite{UzunovaIvUnpublished}. Indeed, the total flux $\Phi$ depends on the value of magnetization far from the vortex center, $\theta_0(\infty)$. For the planar vortices $\theta_0(\infty)=\pi/2$ and the total flux is the integer multiple of the flux quantum. The same situation is with skyrmions, which are characterised by $\theta_0(\infty)=\pi$. In both cases $\Phi$ is an integer multiple of $\Phi_0$ and the  Aharonov-Bohm type of splitting does not appear.}

\cl{In what follows we are considering vortices with the lowest energy: only vortices with $q=1$ are stabilized by the magnetostatic dipole interaction. We expect that for higher $q$ the situation will be qualitatively the same.}

\cl{The structure of magnetic vortex discussed above is basically similar for disk- and ring-shaped particles, but topological properties have peculiarities. For both cases the main topological invariant is the $\pi_1$-topological charge, because the vortex structure is essential for magnons to be scattered on its background. The $\pi_2$-topological charge is an auxiliary quantity, determined by area of part of the sphere, in which the distribution of magnetization $\mathbf{M}$ is mapped. In the disk-shaped particle this mapping covers half of the sphere for planar vortex, and the part other than half for the cone vortex. Contrary, a central part of the vortex, including the vortex core, is absent in the ring, and the topological arguments about $\pi_2$-topological charge are not applicable in this case, in particular, the polarization $p$ is undefined. But as we will see below, this peculiarities does not change the essence of the effect we are investigating.}

\cl{Indeed, in the context of the Aharonov-Bohm effect the most fundamental point of the topological analysis is the introduction of the vector-potential, \cite{aharonovSignificanceElectromagneticPotentials1959}. In the magnon analogue of the effect the vector-potential $\bf{A}$ induced by the vortex topology can be naturally identified from the Landau-Lifshitz equations, if they are transformed to a Schrödinger-like form. We show in the next Section that it can be done for a model that takes into account both local exchange and non-local magnetostatic interactions.}
\cl{We show that the vector-potential induced by the vortex topology results a giant magnon splitting in a ring-shaped magnetic particle, for which a vortex core is absent. This result corresponds exactly to the letter and spirit of the Aharonov-Bohm effect, whose goal was to show the primacy and physicality of the vector potential.}

\section{Theoretical description of magnon modes in the ferromagnetic ring}\label{sec:analytical_model}

In this section we analyze the magnon excitations in the ferromagnetic ring on the background of the cone-state vortex.
 The energy of the soft magnetic ferromagnet, like permalloy, contains the contributions from the isotropic
exchange interaction and the magnetic dipole interaction
\begin{eqnarray}\label{W}
 W=\int d\bm{r}\left[\frac{A}{2 M_s^2}\sum_i{(\partial_i\bm{M})}^2-\bm{M}( \bm{H}_0+\frac{1}{2}
 \bm{H}_m
)\right]
\end{eqnarray}
where $A$ is the exchange constant,
${\bm{H}}_0=4\pi M_s h \bm{e}_z$ is external field.
The  field $ \bm{H}_m$  is determined by the
magnetostatic equations $\textmd{div} (\bm{H}_m + 4\pi
\bm{M}) = 0$ and $\textmd{rot}\ \bm{H}_m = 0$ with the
standard boundary conditions: the continuity of the normal component
of $(\bm{H}_m + 4\pi \bm{M})$ and the tangential component
of $\bm{H}_m$ on the border of the sample. Formally, the sources of
$\bm{H}_m$ can be considered as “magnetic charges” with the volume and surface
charge density being $\rho_\textmd{m}=\textmd{div}\bf{M}$ and
surface charge density is equal to $-(\bm{M} \cdot\bm{n})$,
where $\bm{n}$ is the unit vector normal to the border. The magnetostatic field is separated into volume
$\boldsymbol{H}_\textmd{m}^{\textmd{vol}}$
 and surface contributions including lateral surfaces
 ${\bm{H}}_\textmd{m}^{\textmd{edge}}$
 and ring faces $\boldsymbol{H}_\textmd{m}^{\textmd{face}}$.
 
 In the most general case, the solution of
magnetostatic equations is a Coulomb-type integral of the magnetic
charge density that makes the Landau-Lifshitz equations
integro-differential. However, if we are considering a thin ring, for which $L/R_\mathrm{in},L/R_\mathrm{out}$ can be treated as small parameters, the expressions for demagnetizing field can be essentially simplified as following.

The main term arising from the surface charges of two ring faces is
${\bm{H}}_\textmd{m}^{\textmd{face}}=-4\pi M_z \bm{e}_z$, that corresponds to the energy density of $2\pi
M_z^2$ and provides an effective planar anisotropy. The contribution of the ring edges, ${\bm{H}}_\textmd{m}^{\textmd{edge}}$,  can be taken into account by virtue of boundary condition on $M_r(R)$, see \cite{ivanovMagnonModesThin2002, Guslienko2010}.
The volume  magnetic charges produce the non-local field ${\bm{H}}_\textmd{m}^{\textmd{vol}}=-\nabla \mathcal{F}_{\textmd{in}}$, where the magnetostatic potential in the main order can be
obtained in the from
\begin{eqnarray}\label{Fin}
\mathcal{F}_{\textmd{in}}=-\frac{2\pi L}{\bar{k}}\textmd{div}_\bot
\bm{M}.
\end{eqnarray}
Here $\bar{k}$ takes the eigenvalues of the Laplace operator and determines the decrease in the demagnetizing field in free space outside the ring, for details see App.~\ref{AppA}.

The magnon modes on the vortex background are
investigated considering small deviations $\vartheta, \mu$ from the
vortex ground state
$\phi=\chi+\pi/2+(\sin\theta_0)^{-1}\mu,\;\theta=\theta_0(r)+\vartheta$.
The additional factor $(\sin\theta_0)^{-1}$ is introduced to express
corrections to magnetization $\bm{m}_0=\cos\theta_0{\bf
e}_z+\sin\theta_0\bm{e}_\chi$ in the  form $\delta{\bf
 m}=-\mu\bm{e}_r +\vartheta\cos\theta_0{\bf
e}_\chi-\vartheta\sin\theta_0\bm{e}_z$.
The linearized
Landau-Lifshitz equations in terms of $\vartheta$ and $\mu$ can then be
written as
\begin{eqnarray}\label{munu}
-\frac{1}{\omega_m}\frac{\partial{\mu}}{\partial t}=-l_0^2 \triangle\vartheta+V_1\vartheta
+\frac{2l_0^2\cos\theta_0}{r^2}\frac{\partial\mu}{\partial\chi}+\\
+\frac{L}{2\bar{k}}\frac{\cos\theta_0}{r}\frac{\partial}{\partial\chi}\left(\frac{\partial\mu}{\partial
r}+\frac{\mu}{r}-\frac{\cos\theta_0}{r}\frac{\partial\vartheta}{\partial\chi}\right),
\notag\\
\frac{1}{\omega_m}\frac{\partial{\vartheta}}{\partial t}=-l_0^2 \triangle\mu+V_2\mu
-\frac{2l_0^2\cos\theta_0}{r^2}\frac{\partial\vartheta}{\partial\chi}-\notag\\
-\frac{L}{2\bar{k}}\frac{\partial}{\partial
r}\left(\frac{\partial\mu}{\partial
r}+\frac{\mu}{r}-\frac{\cos\theta_0}{r}\frac{\partial\vartheta}{\partial\chi}\right).\notag
\end{eqnarray}
where $\omega_m=4\pi M_s\gamma$, $\gamma=g\mu_B/\hbar$ is the gyromagnetic ratio, $g\approx 2$, $\mu_B$ is the Bohr magneton, $l_0^2=A/4\pi M_s^2$,
 and the
potentials are
\begin{eqnarray}
V_1&=&\cos 2\theta_0 \left(\frac{l_0^2}{r^2}-1
\right)+h\cos\theta_0, \\
V_2&=& -l_0^2 \left(\frac{d\theta_0}{dr}\right)^2+\cos^2\theta_0 \left(\frac{l_0^2}{r^2}-1
\right)+h\cos\theta_0.\nonumber
\end{eqnarray}

For the dynamics of magnons in the form of plane waves considered far from the vortex core, Eqs.~(\ref{munu}) give the magnon dispersion law as a function of the wave vector $k$, namely
\begin{eqnarray}\label{dl}
\omega={\omega_mk\sqrt{1-h^2+l_0^2k^2}}\sqrt{\left.{l_0^2+\frac{L}{2\bar{k}}}\right.}.
\end{eqnarray}
It takes into account both exchange and magnetostatic interactions.

The non-local dipolar
 interaction  greatly complicates the dynamic
equations and they cannot be reduced neither to Schr\"{o}dinger nor
to generalized Schr\"{o}dinger equations of the form~(\ref{GShr}). Thus Eqs.~(\ref{munu}) rewritten as one equation for single complex-valued
function $\psi=\vartheta+i\mu$ are of the form
\begin{eqnarray}\label{Shr}
i\frac{\partial\psi}{\partial t} =
\hat{H}_1\psi+\hat{H}_2\psi^* +i \omega_m
\left[2l_0^2+\frac{L}{2\bar{k}}\right]\frac{A_\chi}{r}
\frac{\partial\psi}{\partial\chi},
\end{eqnarray}
 where $\hat{H}_1$ and $\hat{H}_2$ are highly anisotropic second order
differential operators.
Here $A_\chi$ can be interpreted as $\chi$-component of the effective topology-induced vector potential
\begin{eqnarray}
\bm{A}(r)=-\frac{ \cos\theta_0}{r}\bm{e}_\chi.
\end{eqnarray}

\cl{Let's now calculate the flux. As we discussed in the previous Section, the vortex structure is characterised by non-zero vorticity $q$, here we take $q=1$, and has a singularity in the center, at point $r=0$. For a disk-shaped particle, the direction of magnetization at point $r=0$ determines polarisation $p=\cos{\theta_0}(0)$, and the total flux can be calculatet in terms of $q$ and $p$. For the ring, there is no magnetization in the central region; however, this central singularity can not be ignored. At the inner radius there is a jump of the vector-potential. Thus, the correct definition of $\bm{A}$ can be written as follows $\bm{A}(r)\Theta(r-R_{\mathrm{in}})$, where $\Theta(r)$ is the Heaviside step function. Then $\textrm{rot}\bm{A}=\mathbf{B}=B\bm{e}_z$, where the magnitude of the field $\mathbf{B}$ is obtained as}
\cl{
\begin{eqnarray}
B= \frac{\cos\theta_0}{r}\delta(r-R_{\mathrm{in}})+\frac{\sin\theta_0}{r}\left(\frac{d\theta_0}{dr}\right)\Theta(r-R_{\mathrm{in}}).
\end{eqnarray}}
\cl{The flux $\Phi=\int \mathbf{B}d\mathbf{s}$ has the form}

\cl{
\begin{eqnarray}
\frac{\Phi}{\Phi_0}=\cos\theta_0(R_{\mathrm{in}})+\int_{\theta_0(R_{\mathrm{in}})}^{\theta_0(\infty)}\sin\theta_0d\theta_0=\cos\theta_0(\infty).
\end{eqnarray}}
\cl{For the cone vortex state created by the external field $\cos\theta_0(\infty)=h$ and the final expression for the total flux is $\Phi=\Phi_0 h$. This result differs from the one obtained for the vortex in the disk, Eq.~(\ref{Phi}). In particular, for the planar vortex in a ring-shaped particle $\Phi=0$ while for the disk case $\Phi=\Phi_0$. But the general properties of the vortex are the same; in both cases there is no Aharonov-Bohm effect for a planar vortex.}

Considering the case when the inner radius of the ring $R_\mathrm{in}$ exceeds the size of the vortex core, it is possible to separate radial and azimuthal variables in Eqs.~(\ref{munu}). Then the wave function of magnon with azimuthal number $m$ takes the form
$\psi=f(r)\cos(m\chi-\omega t)+i g(r)\sin(m\chi-\omega t)$  and operators from Eq.~(\ref{Shr}) are
\begin{eqnarray}\label{HH}
 \hat{H}_1=
 \frac{\omega_m}{2}\left[1-h^2
-2l_0^2\triangle-\frac{L}{2\bar{k}}\triangle_r
 \right],\\
 \hat{H}_2=
 \frac{\omega_m}{2}\left[1-h^2+
\frac{L}{2\bar{k}}\triangle_r
 \right],\notag
\end{eqnarray}
where $\triangle_r={\partial^2}/{\partial
r^2}+(1/r){\partial}/{\partial r }-{1}/{r^2}$. Both
operators $ \hat{H}_1$ and  $ \hat{H}_2$ have terms $1-h^2$
finite at $r\rightarrow\infty$ what makes the $\text{Im} \psi$  the
master function and $\text{Re}\psi$ the slave one.

Then the equation (\ref{Shr}) in the lowest powers of small parameters $L/R_\mathrm{in}$,  $L/R_\mathrm{out}$
and $l_0^2/r^2$ reduces to the Bessel equation and has a general solution expressed in terms of cylindrical functions of the index
\begin{eqnarray}\label{nu}
\nu=\sqrt{\frac{L+2\bar{k}m^2l_0^2}{L+2\bar{k}l_0^2}+\frac{2L+8\bar{k}l_0^2}{L+2\bar{k}l_0^2}
\frac{\omega m h }{\omega_m(1-h^2)}}.
\end{eqnarray}
Therefore, the magnon wave function is expressed in terms of the Bessel functions $J_\nu(kr)$ and $J_{-\nu}(kr)$ of the non-integer, generally speaking transcendental, indexes.

Spectrum of magnons can be calculated on the base of the dispersion law~(\ref{dl}) by substituting the values of $k$ and $\bar{k}$. The wave number $k$ is obtained from the boundary condition on the ring edges and the eigenvalue of Laplace operator $\bar{k}$ is obtained from the condition of equality of the magnetostatic potential inside and outside the magnetic ring. The frequencies found within this analytical approach are computed by the iterative procedure of consistently solving of the boundary conditions. The results are shown by the solid lines in Figs.~\ref{fig:FIG4},~\ref{fig:FIG5} and are discussed in the next section. Detailed description of the procedure is presented in App.~\ref{AppA}. 

Before performing these calculations, some conclusions can be made from the Eq.~(\ref{nu}) for the index of the Bessel function. In general the index $\nu$ is not-integer even for the planar vortex, and the reason is the competition of magnetostatic and exchange interactions. The limit case of exchange domination corresponds to large $m$, $m^2>m_c^2=L/(\bar k l^2_0)$, $\nu = |m|$, as for local easy-plane model \cite{ivanovMagnonModesMagnonvortex1998}, giving the rise for growing of all the frequencies with increasing of $|m|$. For small $|m|<m_c$, the magnetostatic interaction dominated and $\nu \sim 1$. It is giving an anomalous dispersion, decreasing of frequencies for growing $|m|$ \cite{buessExcitationsNegativeDispersion2005, guslienkoDynamicOriginAzimuthal2008}. For planar vortices, $h=0$, $\nu$ depends on $|m|$ only and the doublets with $m=\pm |m|$ are not splitted. For the cone vortex state, the situation changes drastically, $\nu(m) \neq \nu(-m)$, that leads to the doublet splitting proportional to $h$. Note that the same mathematical features, namely presence of Bessel functions with non-integer index, which depends on the sign of azimuthal quantum number of electronic states, appear in the standard Aharonov-Bohm problem \cite{aharonovSignificanceElectromagneticPotentials1959}.

\section{Comparison with numerical simulations}\label{sec:validation}

\subsection{Micromagnetic modeling}

To confirm our theoretical predictions, we performed micromagnetic modeling using an in-house-developed dynamic-matrix method similar to the one by Grimsditch \cite{grimsditchMagneticNormalModes2004} or d`Aquino  \cite{daquinoNovelFormulationNumerical2009,daquinoComputationMagnetizationNormal2012} \color{black}. We start with the calculation of the nonlinear ground state magnetization $\mathbf{m}_0$, that determines the vortex structure. For this we are solving static Landau-Lifshitz equation $[\mathbf{m}_0\times\mathbf{H}_{\text{eff}}]=0$. Then to calculate the frequencies of the magnon eigenmodes, we are solving linearized dynamic Landau-Lifshitz equation, 
\begin{equation}\label{eq:linllg}
    \frac{\partial{\bm{m}}}{\partial t}=-\gamma\mu_0  \big\{\bm{m}\times \bm{H}_{\text{eff}}[\bm{m}_0] + \bm{m}_0\times \delta \bm{h}[\delta\bm{m}]\big\}.
\end{equation}
which describes the temporal evolution of the magnetization dynamics for small variations $\delta\bm{m}$ around equilibrium distribution $\bm{m}_0$. The static effective field $\bm{H}_{\text{eff}}$ contains Zeemann, exchange and dipolar interactions, while its dynamic counterpart $\delta \bm{m}$ contains only the latter two (as the external field is taken as constant in time).
To proceed, Eq.~\eqref{eq:linllg}\color{black}~is expanded in terms of linear magnon modes $\delta\bm{m}=\sum_\nu \bm{m}_j \exp(i\omega_j t) + \mathrm{c.c.}$ with  mode index $j$, eigenfrequencies $\omega_j$ and spatial mode profiles $\bm{m}_j(\bm{r})$. With this, the linearized equation can be written as an eigenvalue problem 
\begin{equation}
\frac{\omega_j}{\gamma \mu_0 M_\mathrm{s}} \bm{m}_j = \hat{\bm{D}} \bm{m}_j
\end{equation}
for the resulting dynamic matrix $ \hat{\bm{D}}$ which is, subsequently, discretized using a finite-element method. The dipolar fields generated by the static and dynamic magnetization are calculated using the hybrid finite-element/boundary-element method by Fredkin and Koehler \cite{fredkinHybridMethodComputing1990}. The frequencies $\omega_j$ and spatial mode profiles $\bm{m}_j(\bm{r})$ are then obtained by diagonalizing the dynamic matrix using an iterative Arnoldi-Lanczos method \cite{lanczosIterationMethodSolution1950,arnoldi1951principle}. Details of the used dynamic-matrix method are also found in Ref.~\cite{korberFiniteelementDynamicmatrixApproach2021}. We note that such an approach directly yields the exact oscillation frequencies (up to numerical errors). 

For our calculations, typical material parameters of the softmagnetic alloy Ni$_{80}$Fe$_{20}$ were used. In particular, we took a saturation magnetization of $M_\mathrm{s}=\SI{796}{\kilo\ampere/m}$, an exchange stiffness constant of $A_\mathrm{ex}=\SI{13}{\pico\joule/m}$ and a reduced gyromagnetic ratio of $\gamma/2\pi= \SI{28}{\giga\hertz/\tesla}$. Three different magnetic rings with an inner radius of $R_\mathrm{in}=\SI{50}{\nano\meter}$, a thickness of $L=\SI{20}{\nano\meter}$ and a varying outer radius $R_\mathrm{out}=250$, $350$ and $\SI{450}{\nano\meter}$ were discretized using small tetrahedrons with an average edge length of \SI{5}{\nano\meter}. The magnetization of the ring was brought into a cone-state vortex using a static out-of-plane magnetic field. For each value of the field, the magnetization was first initialized in a vortex state and then subsequently relaxed into its equilibrium state by minimizing the total magnetic energy. Finally, the lowest 100 eigenmodes for each field (500 for zero field) were obtained using our finite-element dynamic-matrix code. The azimuthal indices $m$ of the modes were obtained automatically from the spatial mode profiles using a cylindrical discrete Fourier transform.

\subsection{Alternating dispersion at zero field}

At first, we compare theory and numerical calculations for the case of zero applied field, where the rings are magnetized completely in-plane, and they are in a circular vortex state. In this case, the Aharonov-Bohm flux is equal to an integer number of effective flux quanta, which means no Aharonov-Bohm splitting and, therefore, the modes with opposite azimutal index $\pm m$ form degenerate doublets. Fig.~\ref{fig:FIG4} shows the frequencies $\omega/2\pi$ of the modes with different radial indexes $n=1,2,...$ in a vortex ring for two different outer radii as a function of the modulus of the azimuthal indices $\abs{m}$.


\begin{figure}[h]
    \centering
    \includegraphics{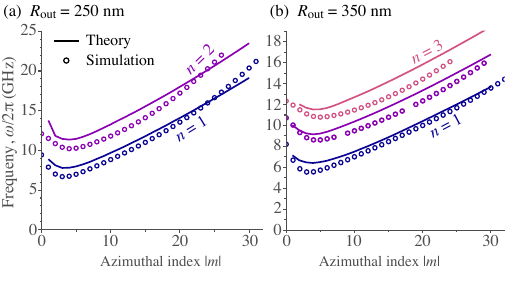}
    \caption{Comparison of dispersions obtained by our analytical model (lines) and numerical calculations (points) for rings with different outer radii, (a) \SI{250}{nm} and (b) \SI{350}{nm}, in the absence of external magnetic field. The frequencies of the modes with different radial index $n$ are plotted in function of the modulus of the azimuthal mode index $\abs{m}$.}
    \label{fig:FIG4}
\end{figure}

We observe a good agreement between our analytical model and numerical calculations. In particular, the alternating character of the dispersion with increasing $\abs{m}$, which arises from the balance between dipolar and exchange interactions, is recovered. The analytical description compares the exchange length $l_0$ and the thickness of the sample $L$ with all other spatial parameters of the system, including the spatial variation of the magnetization. In this small $m$ regime, the dipolar interaction is dominant and favours closure of the magnetic flux by bringing dynamic magnetic surface charges closer together (by increasing $\abs{m}$). For small azimuthal indices the dispersion is negative, {i.e.} the frequencies decrease with increasing $\abs{m}$. This effect for magnons propagating parallel to the equilibrium magnetization is known as backward-volume behavior in thin magnetic films and has been measured in vortex-state magnetic nanodots for the first time by Buess \textit{et al.} in Ref.~\cite{buessExcitationsNegativeDispersion2005}. As $m$ increases, the characteristic size of spatial variation of magnetization $2\pi R_\mathrm{out}/m$ becomes comparable to the sample thickness. For large $\abs{m}$, the exchange interaction is dominant and leads to an increase in frequency with increasing azimuthal index proportional to $m^2$.

In the intermediate region, a minimum of the dispersion is present, which shifts to larger $\abs{m}$ as the outer radius of the ring increases and, thus, dipolar fields become more important. For even larger radii, it is already well-known that this negative dispersion and its shift towards larger radii can lead to resonant three-magnon decay of the second- or even first-order radial modes $n=1,2$, which, in return, allows for the experimental observation of azimuthal modes with very large $\abs{m}$
\cite{schultheissExcitationWhisperingGallery}.

\subsection{Doublet split}

When an external magnetic field $H_0$ is present, the degeneracy of doublets with different signs of the azimuthal number $m$ is removed and we can observe a splitting of doublets. 

This splitting is the topological analogue of the Aharonov-Bohm effect for magnons scattered by the core of magnetic vortex, despite the fact that the ring-shaped sample does not have the vortex core region. Indeed, in the cone state of the vortex provided with an external magnetic field, the total flux $\Phi$ defined by formula (\ref{Phi}) is not an integer of a flux quantum. The magnon wave functions are proportional to Bessel functions of the index $\nu$, given by Eq.~(\ref{nu}), which is dependent on sign of the azimuthal number $m$. Therefore  modes with different sense of gyration have different frequencies.
The existence of the doublet splitting is confirmed by the simulations.

The frequencies as a functions of external out-of-plane field $\omega(H_0)$, showing the doublet splitting, that are obtained from both theory and simulations are presented in Fig.~\ref{fig:FIG5}. To make the comparison more clear we introduce a correction shift $C$, representing the frequency at zero field, $C=\omega{(0)}/2\pi$. In Fig.~\ref{fig:FIG5} we are using $C$ as a single adjustable parameter of the theory. The calculated values of $\omega(0)$ are presented in Fig.~\ref{fig:FIG4}, where can be seen that the difference between the frequency values obtained from numerical calculations and analytics is not more than $10\%$. \cl{It is quite a good agreement, taking into account the simplifications we made.} {The magnitude of the splitting is well reproduced by the theory without any adjustable parameters, because the splitting is robust due to its topological nature.} 

\begin{figure}[h!]
    \centering
    \includegraphics{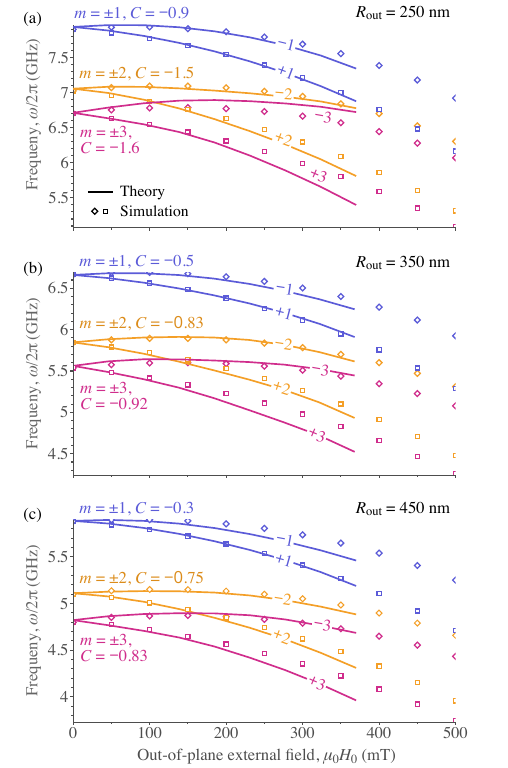}
    \caption{ Doublet split: the frequency evolution of the first three azimuthal modes calculated analytically (lines) and numerically (points) for rings with different outer radii, (a) \SI{250}{nm}, (b) \SI{350}{nm} and (c) \SI{450}{nm}, in function of the out-of-plane static magnetic field strength. Correction shift $C$ is zero-field frequency, $ C=\omega(0)/2\pi$, which we adjust to compare the results.}
    \label{fig:FIG5}
\end{figure}

\section{Conclusion}

In this work, we have focused on studying how the non-trivial topology of a magnetic dot in the form of a ring affects the dynamics of magnons. An exact analytical calculation in such a geometry is not possible, due to the presence of a three-dimensional non-local field. Therefore, when constructing the theory, it was necessary to make an approximation where the thickness of the sample is the smallest spatial parameter of the problem. 

{The developed approximate theory, which contains both local exchange and non-local dipolar interactions can describe the} nontrivial behavior of dependence of the mode frequency on azimuthal mode number $m$. The analytical results are confirmed by numerical calculations, using a finite-element dynamic-matrix approach.

We have also shown that a nonzero external magnetic field, applied perpendicular to the ring plane, leads to the splitting of doublets with the opposite sign of azimuthal mode number. This phenomenon is a consequence of the interaction of magnons with the cone vortex state. In this sense, it is a topological analogue of the Aharonov-Bohm effect, when splitting occurs at a non-integer number of flux quanta.
The numerical data for the splitting are reproduced by theory using a single adjustable parameter, namely the frequency at zero field.
Importantly, the presence of the splitting itself is a topologically determined consequence of the nontrivial Aharonov-Bohm effect. 

\color{black}
Apart from the fundamental implications of this study, the described doublet splitting can have various consequences for different applications based on magnons. For example, magnetic rings in the vortex state can be used as magnon filters or nonlinear resonators \cite{wang2020nonlinear}. We expect the behavior of such devices to change under the application of a static out-of-plane field. As another example, the magnons in vortex-state disks or rings can undergo multi-magnon scattering processes which adhere to selection rules that depend on the spatial profiles of the modes involved \cite{schultheissExcitationWhisperingGallery}. These selection rules lead to scattering within well-defined channels that can be exploited to achieve magnon-based pattern recognition \cite{korberPatternRecognitionReciprocal2023}. The Aharonov-Bohm flux in the cone-vortex state can potentially modify such selection rules, as it influences not only the mode frequencies but also their spatial profiles. Thus, magnetic structures in the cone-vortex state also provide an interesting system to study in the interplay between topological properties, nonlinear dynamics and applications thereof. 
\color{black}

\section*{Acknowledgements}
L.K. and A.K. gratefully acknowledge financial support by the Deutsche Forschungsgemeinschaft within the programs KA 5069/1-1 and KA 5069/3-1.

\section{Appendix A \label{AppA}}
We consider the case where the thickness of the ring, $L$, is the smallest size of the system. This allows us to make some approximate analytical calculations for dynamic equations that are quite complex in the general case.

Taking the ratios $L/R_\mathrm{in}$ and $L/R_\mathrm{out}$ as small parameters, the main contribution of the surface charges of the two faces of the ring we can find directly from the continuity of the component
of $(\bm{H}_m + 4\pi \bm{M})$ along $z$ direction, which immediately gives
${\bm{H}}_\textmd{m}^{\textmd{face}}=-4\pi M_z \bm{e}_z$. In
the expression for the energy density this corresponds to $2\pi
M_z^2$ and provides an effective easy-plane anisotropy. The term
${\bm{H}}_\textmd{m}^{\textmd{edge}}$  can be taken into account
by virtue of boundary condition on $M_r(R)$ on the ring edges, from which the magnon
wave number is determined. Thus ${\bm{H}}_\textmd{m}^{\textmd{edge}}$  do not contribute to the Landau-Lifshitz
equations, for details see ref. \cite{ivanovMagnonModesThin2002,Guslienko2010}.

The volume  magnetic charges, ${\bf
H}_\textmd{m}^{\textmd{vol}}=-\nabla \mathcal{F}_{\textmd{in}}$, are the sources of a non-local field, that cannot be computed analytically in the general case. 
The corresponding magnetostatic potential $\mathcal{F}_{\textmd{in}}$ satisfies the Poisson's equation
$\triangle\mathcal{F}_{\textmd{in}}=4\pi \rho_\textmd{m}$ and is
continuous on the ring faces. For the thin ring under consideration the volume
magnetic charge density can be assumed effectively two
dimensional: $\rho_\textmd{m}(\bm{r})=L\delta
(z)\textmd{div}_\bot\bm{M}$, where
\begin{eqnarray}
\textmd{div}_\bot\bm{M}=\left(\nabla_{\bot}\cdot\bm{M}\right)=\frac{\partial(rM_r)}{r\partial
r}+\frac{\partial(M_\chi)}{r\partial
 \chi}.
\end{eqnarray}
 Here $\bm{M}=(M_r,M_\chi,M_z)$ are the magnetization components in cylindrical coordinates.
Then the potential in the outer region $\mathcal{F}\left(|z|>{L}/{2}\right)=\mathcal{F}_{\textmd{out}}$ can be written as the general solution of Laplace equation, $\triangle\mathcal{F}_{out}=0$, that is finite at
the origin and decaying at infinity. Namely,  $\mathcal{F}_{\textmd{out}}=\sum_mJ_m(\bar{k}r)\xi(\chi,t)e^{-\bar{k}|z|}$, where $\xi(\chi,t)$ is
a harmonic function of the $m\chi-\omega t$. For the thin magnetic dot the potential in the inner
region $\mathcal{F}\left(|z|<{L}/{2}\right)=\mathcal{F}_{\textmd{in}}$ can be
assumed homogeneous in $z$-coordinate. It allows to separate $z$- and
in-plane variables in the Poisson's equation and
integrate it over $z$ throughout the space. 
Herewith $z$-derivative is explicitly singled out in the Laplace operator, $\triangle={\partial^2}/{\partial z^2}+\triangle_\bot$. 
Since there are no magnetic charges in the outer region the only nonzero contributions in the
integral from $\mathcal{F}_{\textmd{out}}$ are the contributions from the regions in the close
vicinity of the two faces of the ring
\begin{eqnarray}
\left.\frac{\partial}{\partial
z}\mathcal{F}_{\textmd{out}}\right|^{{L}/{2}+0}_{-{L}/{2}-0}+L\triangle_\bot\mathcal{F}_{\textmd{in}}=4\pi
L\textmd{div}_\bot\bm{M}.
\end{eqnarray}
Taking into account the equality of potentials on faces, the first
term is $-2\bar{k}\mathcal{F}_{\textmd{in}}$ and in the main order we obtain for the potential inside the sample
\begin{eqnarray}\label{Fin}
\mathcal{F}_{\textmd{in}}=-\frac{2\pi L}{\bar{k}}\textmd{div}_\bot
\bm{M}.
\end{eqnarray}
 Note, that magneto-static potential is nonzero only for deviations from the vortex ground state. 
The wave number $k$ is obtained from the boundary condition on the on the ring edges. 
The eigennumber of Laplace operator $\bar{k}$ should be found from continuity of the potential $\mathcal{F}$ and its tangential field $-\nabla_{\bot}\mathcal{F}$,
\begin{eqnarray}\label{kbc}
\left.\mathcal{F}_{\textmd{out}}\right|_{|z|=\frac{L}{2}}=\left.\mathcal{F}_{\textmd{in}}\right|_{|z|=\frac{L}{2}},\\
\left.\nabla_{\bot}\mathcal{F}_{\textmd{out}}\right|_{|z|=\frac{L}{2}}=\left.\nabla_{\bot}\mathcal{F}_{\textmd{in}}\right|_{|z|=\frac{L}{2}}. \notag
\end{eqnarray}

Separating the variables into radial and angular components and also using the explicit form of the wave function, we can solve (approximately) Eqs. (\ref{kbc}).

In the main approximation the magnon wave function $\psi$  has the form
\begin{eqnarray}\label{ef}
\psi(r,\chi)=\left[ J_\nu(kr)-\sigma_mJ_{-\nu}(kr)\right]\sin(m\chi-\omega t),
\end{eqnarray}
where $\sigma_m={J_\nu(kR_\mathrm{in})}/{J_{-\nu}(kR_\mathrm{in})}$ and noninteger index $\nu$ of Bessel functions $J_\nu$ and $J_{-\nu}$  is given by Eq.~(\ref{nu}). Here is taken into account that the $r$-part of the edge conditions on the outer and inner radii of the ring can be considered fixed without significant loss of accuracy, $g(R_\mathrm{in})=g(R_\mathrm{out})=0$. Whereas the angular part is naturally satisfied for the harmonic function $\xi(\chi,t)$.

The wave number $k$ in the dispersion law can be computed from the condition on the outer edge, $\psi(R_\mathrm{out},\chi)=0$, and the eigenvalue of Laplace operator $\bar{k}$ is the solution of Eqs.~(\ref{kbc}). Since we are using the thin-ring approximation to introduce the dipole interaction potential, the conditions of equities of potentials and fields (\ref{kbc}) can not be satisfied exactly.
Therefore, we are using for calculations conditions (\ref{kbc}) averaged over the surface of the ring with the magnon eigenfunctions $\psi$, Eq.~(\ref{ef}). The first order of these perturbations when averaged over the surface is equal to zero, since all quantities are proportional to $\sin(m\chi-\omega t)$. The next order of perturbations will give us the equation for calculating $\bar{k}$:
\begin{eqnarray}\label{bc_psi}
&\int_{R_\mathrm{in}}^{R_\mathrm{out}}rdr\int_0^{2\pi}d\chi\left[\mathcal{F}_{in}^2-\mathcal{F}_{out}^2\right]|\psi|^4=0,& \\
&\int_{R_\mathrm{in}}^{R_\mathrm{out}}rdr\int_0^{2\pi}d\chi\left[(\nabla_{\bot}\mathcal{F}_{in})^2-(\nabla_{\bot}\mathcal{F}_{out})^2\right]|\psi|^4=0.&
\nonumber
\end{eqnarray}
Here we need to substitute the explicit expressions for potentials $\mathcal{F}_{in}$ and $\mathcal{F}_{out}$, and for the vawe function $\psi$, Eq.~(\ref{ef}).

In its turn the equation for calculating ${k}$ has the form 
\begin{eqnarray}\label{bc_k}
 J_\nu(kR_\mathrm{out}){J_{-\nu}(kR_\mathrm{in})}={J_\nu(kR_\mathrm{in})}J_{-\nu}(kR_\mathrm{out}).
\end{eqnarray}
Finally, spectrum of magnons is obtained by the iterative calculation of the dispersion law~(\ref{dl}),  Eq.~(\ref{bc_psi}) and Eq.~(\ref{bc_k}).


\end{document}